\documentclass[aps,prx,onecolumn,amsmath,superscriptaddress,amssymb,nofootinbib,11pt]{revtex4-1}
\usepackage{graphicx}
\usepackage{ulem}
\usepackage{soul}
\usepackage[colorlinks, linkcolor=blue]{hyperref}
\usepackage{verbatim}
\usepackage{esint}
\renewcommand{\vec}[1]{\boldsymbol{#1}}
\def \k {{\vec k}}

\def \e {\epsilon}
\def \ve {\varepsilon}
\def \r {{\vec r}}

\def \q {{\vec q}}
\def \Q{{\vec Q}}

\def \ve {\varepsilon}

\def \K {{\vec K}}

\def \n{{\mathbf{n}}}
\def \L{{\cal{L}}}

\def \beq {\begin{eqnarray}}
\def \eeq {\end{eqnarray}}
\def \tn {\textnormal}
\def \PP {{\cal {P}}}
\def \M {{\cal {M}}}

\begin{document}
\title{Density wave instabilities of fractionalized Fermi liquids}
\author{Debanjan Chowdhury}
\affiliation{Department of Physics, Harvard University, Cambridge, Massachusetts-02138, U.S.A.}
\author{Subir Sachdev}
\affiliation{Department of Physics, Harvard University, Cambridge, Massachusetts-02138, U.S.A.}
\affiliation{Perimeter Institute of Theoretical Physics, Waterloo, Ontario-N2L 2Y5, Canada. }
\begin{abstract}
Recent experiments in the underdoped regime of the hole-doped cuprates have found evidence for an incommensurate charge density wave  state. We present an analysis of the charge ordering instabilities in a metal with antiferromagnetic correlations, where the electronic excitations are coupled to the fractionalized excitations of a quantum fluctuating antiferromagnet on the square lattice. The resulting charge density wave state emerging out of such a fractionalized Fermi-liquid (FL*) has wavevectors of the form $(\pm Q_0,0),~(0,\pm Q_0)$, with a predominantly $d$-form factor, in agreement with experiments on a number of different families of the cuprates. In contrast, as previously shown,
the charge density wave instability of a nearly antiferromagnetic metal with a large Fermi surface, interacting via short-range interactions, has wavevectors of the type $(\pm Q_0,\pm Q_0)$. Our results show that the observed charge density wave appears as a low-energy instability of a fractionalized metallic state linked to the proximity to an antiferromagnetic insulator, and the pseudogap regime can be described by such a metal at least over intermediate length and energy scales.
\end{abstract}
\maketitle
\section{Introduction}
\label{intro}
The central puzzle of the hole-doped cuprates at low doping is the origin of the ``pseudogap" --- a suppression in the density of states at the Fermi-level below a temperature, $T^*$--- and its relation to other symmetry-broken states found at lower temperatures. Most notably, recent experiments on a number of different families of the hole-doped cuprates have detected the onset of an incommensurate charge density wave (CDW) state at a temperature, $T_c<T_{\tn{cdw}}<T^*$ which competes with superconductivity below the superconducting $T_c$ \cite{Ghi12,DGH12,SH12,comin13,neto13,DGH13,MHJ11,MHJ13,CP13, JH02,AY04,JSD1011,comin2,SSJSD14}. One of the primary motivations behind investigating the nature of the CDW is the hope that a better understanding of this state would lead to a better understanding of the ``normal" state above $T_c$ out of which it emerges. 

\begin{figure}
\begin{center}
\includegraphics[width=1.0\columnwidth]{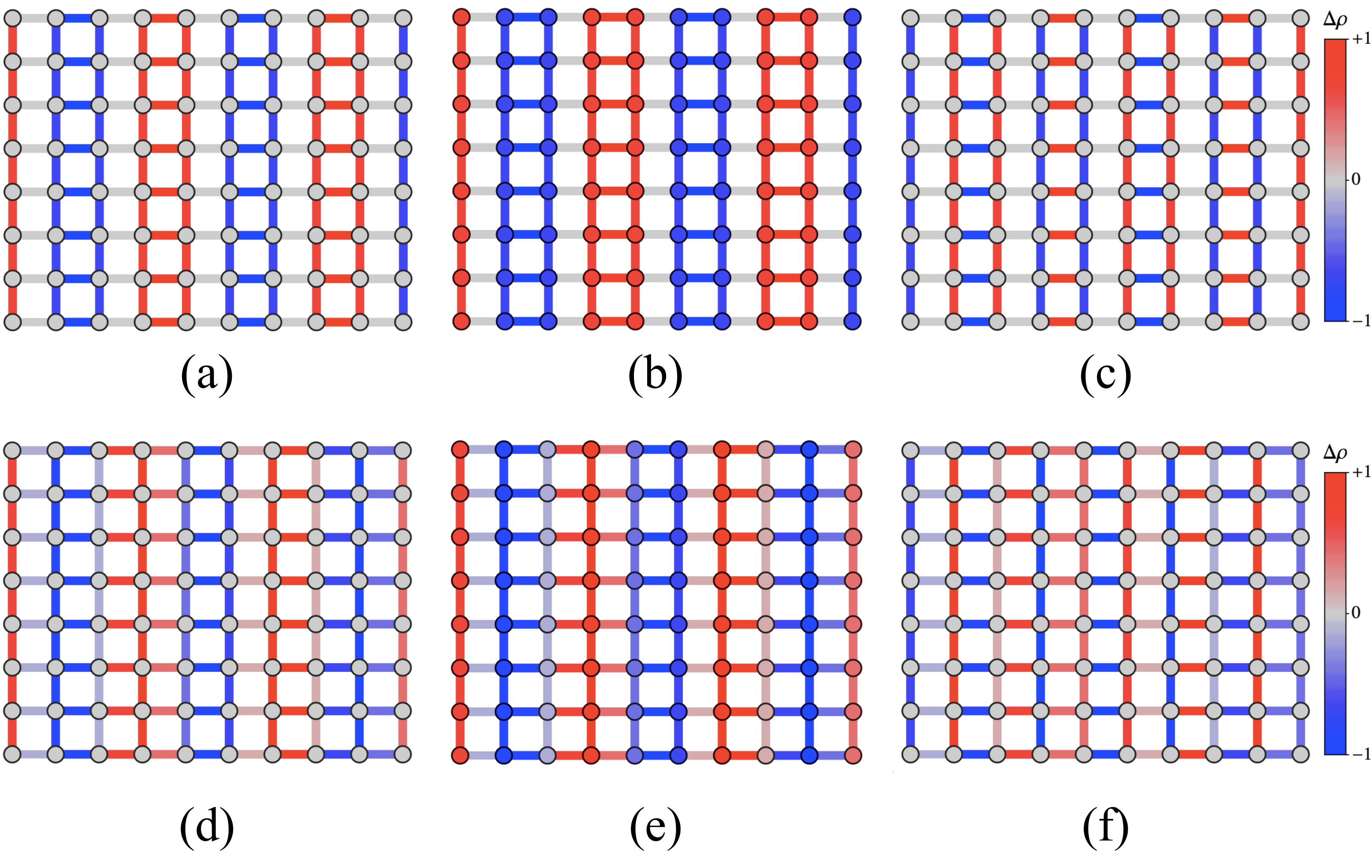}
\end{center}
\caption{Real-space visualization of unidirectional BDW (charge-stripe) with components: $P^s_\Q(\k)=P_s$, $P^{s'}_\Q(\k)=P_{s'}(\cos k_x + \cos k_y)$, and $P^d_\Q(\k)=P_d(\cos k_x - \cos k_y)$. (a)-(c) we plot the charge modulation for a commensurate wavevector, $\Q=2\pi(\frac{1}{4},0)$, while (d)-(f) plot the same quantity for an incommensurate wavevector, $\Q=2\pi(0.3,0)$. The parameters used are: (a), (d) $P_s=0,~P_{s'}=1,~P_d=0$. (b), (e) $P_s=1,~P_{s'}=1,~P_d=0$. (c), (f) $P_s=0,~P_{s'}=0,~P_d=1$. We have included phases in the definitions of $P_{s,s',d}$ in order to make the charge distribution ``bond-centered" for the case of the commensurate wavevector; however, other choices of the phases are also allowed.}
\label{bdw}
\end{figure}

A number of recent theoretical works \cite{MMSS10,metzner,yamase,SSRP13,kee,SSJS14,DHL13,HMKE,HMKE13,HF14,AASS14,AASS14b,DCSS14,YWAC14,norman14,ATAC14,bulut,AKB14,EAK14} have tried to approach this problem from a weak-coupling approach, where the CDW is interpreted as an instability of a large Fermi surface in the presence of strong antiferromagnetic (AFM) exchange interactions. There are two fundamental properties associated with a CDW--- its wavevector, $\Q$, and its form factor, $P_\Q(\k)$, --- where the CDW is expressed in real space as a bond-observable, $P_{ij} = \left\langle c_{i \alpha}^\dagger c_{j \alpha}^{\vphantom\dagger} \right\rangle$ ($c_{i \alpha}$ annihilates an electron on the Cu site $i$ with spin $\alpha$) given by
\beq
P_{ij}=\sum_\Q\bigg[\sum_\k P_\Q(\k) e^{i\k\cdot(\r_i-\r_j)} \bigg]e^{i\Q\cdot(\r_i+\r_j)/2}.
\label{Dij}
\eeq
Conventional charge density waves have only on-site charge modulation with $P_\Q(\k)$ independent of $\k$; this is not the case in the context of the cuprates. The experiments on certain families of the cuprates, such as YBCO, BSCCO and Na-CCOC, have found strong evidence for the wavevector $\Q$ to be of the form: $(\pm Q_0,0)$ and $(0,\pm Q_0)$, where $Q_0$ decreases with increasing hole-doping and is believed to nest portions of a putative large Fermi surface in the vicinity of, but away from, $(\pi,0)$ and $(0,\pi)$. Moreover, recent phase-sensitive STM experiments \cite{SSJSD14} and other X-ray measurements \cite{comin2} have unveiled the form factor $P_\Q(\k)$ to be predominantly of a $d-$wave nature, $i.e.$ $P_\Q(\k)\sim(\cos k_x-\cos k_y)$. Hence, the CDW should be more appropriately referred to as a ``bond density" wave (BDW). Fig.\ref{bdw} provides an illustration of unidirectional BDWs of different types in real space for both commensurate as well as incommensurate wavevectors. 

We note in passing that recent X-ray observations \cite{AADH14} in the La-based cuprates indicate a dominant $s'$-form factor (where $P_\Q(\k)\sim(\cos k_x + \cos k_y)$), and this has been
ascribed to the presence of magnetic `stripe' order in these compounds, and is in agreement with computations in such states \cite{EAK14}.
Further, computations \cite{ATSS14} of density wave instabilities in the
presence of commensurate antiferromagnetism show suppression of the $d$-form factor with increasing magnetic order.

%%%%%%%%%%%%%%%%%%%%%%%%%%%%%%%%%%%%%%%%%%%%%%%%%%%%%%%%%%%
%%%%%%%%%%%%%%%%%%%%%%%%%%%%%%%%%%%%%%%%%%%%%%%%%%%%%%%%%%%

Without magnetic ordering, the form factor does indeed come out to be predominantly $d-$wave within all the weak-coupling approaches \cite{MMSS10,SSRP13,SSJS14,AASS14,AASS14b,DCSS14,YWAC14}. However, the wavevector of the leading density-wave instability within all of these approaches is always of the diagonal type, $(\pm Q_0,\pm Q_0)$. This feature can be traced back to the absence of a pre-existing gap in the anti-nodal region of the large Fermi surface normal state.
It is also related to 
the remnant of an emergent SU(2) symmetry associated with AFM exchange interactions, that maps particles to holes and vice versa \cite{MMSS10,SSMMMP12}; therefore $d-$wave superconductivity, which is the leading instability in the problem, gets mapped to $d-$form factor BDW with the diagonal wavevectors. While the diagonal wavevector is a serious drawback of the weak-coupling approaches, various scenarios have been proposed under which the experimentally observed state might be favored over the diagonal state \cite{AASS14,DCSS14,HMKE13,YWAC14}.

Here we shall explore the consequences of gapping out the anti-nodal region by examining the density wave instabilities of a metallic state
denoted \cite{TSSSMV} the fractionalized Fermi liquid (FL*). An independent related analysis has been carried recently by Zhang and Mei \cite{LZJM14}
using the Yang-Rice-Zhang (YRZ) ansatz \cite{YRZ}
for the fractionalized metal. The FL* has similarities to the YRZ ansatz \cite{YQSS10}, but as we shall review below, can be derived systematically from microscopic models while keeping careful track of its `topological' order \cite{TSSSMV} (see also Ref.~\onlinecite{Wen12}). These density wave instabilities were also discussed in 
Refs.~\onlinecite{comin13,MV12}, but without allowance for non-trivial form factors for the density wave.

FL* phases are most conveniently described within multi-band models, such as in Kondo lattice models for heavy-Fermion systems \cite{TSSSMV} or Emery-type models suited for the cuprates, where the spins in only one of the bands go into a quantum-disordered (spin-liquid) state. In this work, we shall take a different route that has been adopted earlier in Refs.\cite{RK07,RK08,SS09,YQSS10}. This approach is best understood as follows \cite{SSMMMP12}: Imagine that we start from a metal with long-range AFM order, so that the Fermi surface has been reconstructed into hole-pockets.\footnote{In general, electron pockets could also appear, but let us assume that the parameters are such that only the hole-pockets are present.} If we now want to describe the loss of antiferromagnetism, then the transitions where the long-range order is lost and the entire Fermi surface is recovered need not be coincident. In particular, it is possible to have two separate transitions, where at the first transition the {\it orientational} order of the AFM is lost over long distances, while the magnitude of the AFM order remains fixed; the large Fermi surface is recovered at the second transition.\footnote{The description of this transition was studied using a SU(2) gauge-theory \cite{SS09}; there could either be a direct transition from the AFM metal to the large FS described by conventional SDW criticality (see Ref.\cite{MMSS10} and references therein), or, via an intermediate non-Fermi liquid phase with a large FS and gapless SU(2) photons. We shall not focus on any of these transitions here, which have been studied in more details elsewhere \cite{DCSS14c}.} Therefore, it is possible to have an intermediate phase with a locally well-defined AFM order (over distances of the order of a short correlation length, $\xi$). As we will review below, this intermediate phase can have hole pockets. Because there is no broken
symmetry, the Luttinger theorem on the Fermi surface volume is violated, but this is permitted because of the presence of topological order \cite{TSSSMV}.

%%%%%%%%%%%%%%%%%%%%%%%%%%%%%%%%%%%%%%%%%%%%%%%%%%%%%%%%%%%
\begin{figure}
\begin{center}
\includegraphics[width=0.8\columnwidth]{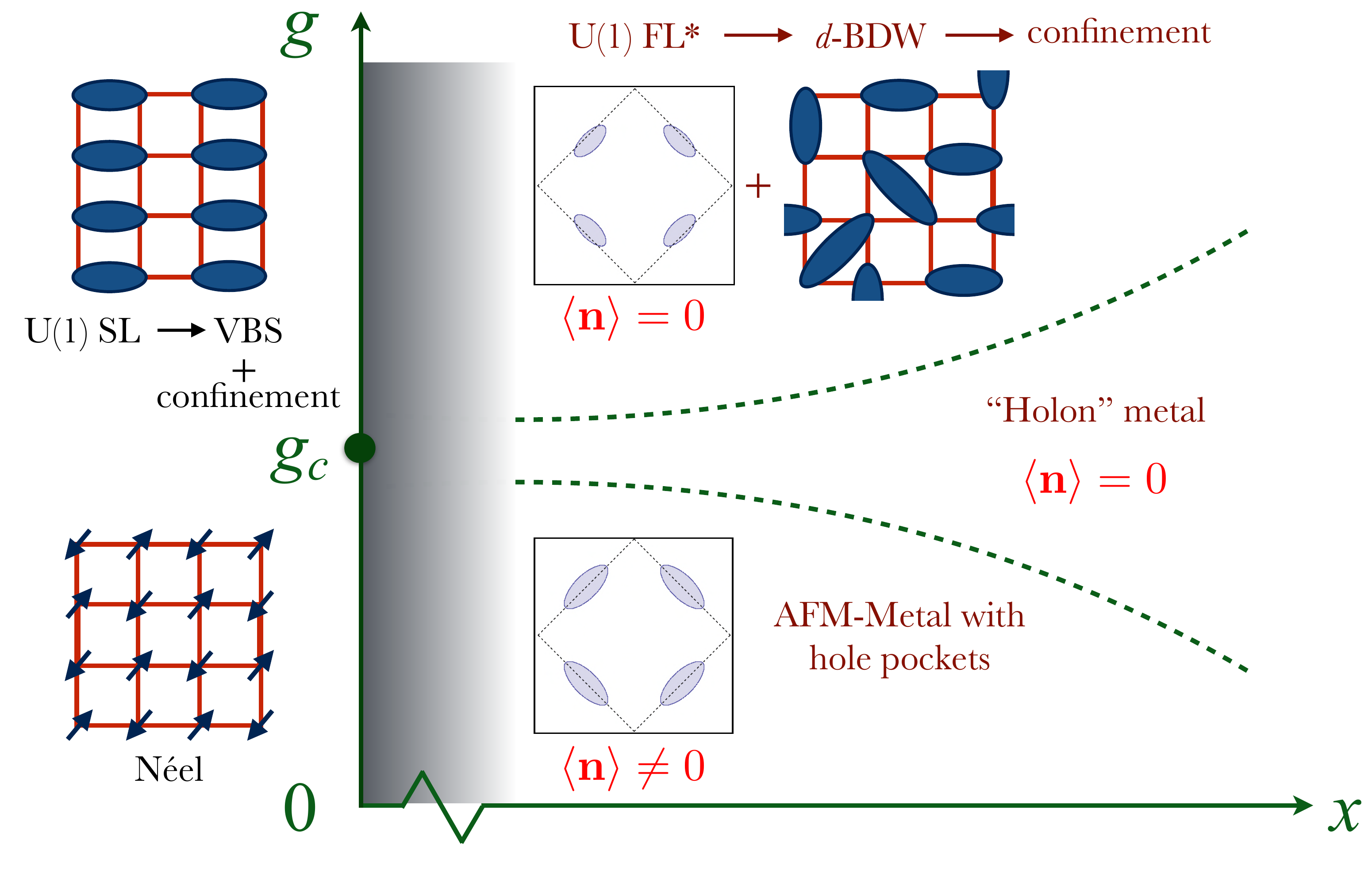}
\end{center}
\caption{Phase diagram of a quantum antiferromagnet as a function of the parameter, $g$, which controls the strength of quantum fluctuations and the carrier density, $x$. The $x=0$ phases are insulators, and we assume that the insulator has a deconfined
quantum critical point \cite{senthil1,senthil2} at a $g=g_c$. The arrows on the labels of the phases indicate crossovers upon going to lower energies and longer distances.
For $x=0$ and $g<g_c$, the ground state is an ordered antiferromagnetic insulator, while for $x=0$ and $g>g_c$ the ground state is a valence bond solid (VBS),  which arises as an instability at long confinement scale of a U(1) spin-liquid (SL). 
The $g=g_c$, deconfined quantum critical point, broadens into a ``holon" metal phase 
above a small, finite $x$. Binding of the emergent spinon and holon excitations in the holon metal phase leads to the FL* phase shown for $g>g_c$ and non-zero $x$. We do not discuss the small $x$ (shaded grey) region in this paper \cite{RK07}. The BDW instabilities of the U(1) FL* have a $d-$form factor and could be relevant for the observations in the non La-based cuprates. The physical trajectory in the phase diagram (for the non-La cuprates) corresponds to not only changing $x$, but also $g$, {\it i.e.\/} the trajectory runs diagonally. The eventual confinement transition out of the $d-$BDW state must occur, and its description remains an important topic for future work.  }
\label{phases}
\end{figure}
%%%%%%%%%%%%%%%%%%%%%%%%%%%%%%%%%%%%%%%%%%%%%%%%%%%%%%%%%%%

In the context of one-band models linked to the physics of the cuprates, our realization of the FL* state is linked to the analysis presented in Ref.~\onlinecite{RK07}.
We used their and subsequent results \cite{RK08,SS09,YQSS10,EGMSS11,MPSS12} to obtain the global phase diagram presented in Fig.~\ref{phases}.
The starting point of this phase diagram is a deconfined quantum critical point \cite{senthil1,senthil2} in the antiferromagnet at zero doping, while tuning a coupling constant $g$, which measures the degree of ``frustration" in the antiferromagnet.
This critical point is described by a deconfined gauge theory involving a U(1) gauge field $A_\mu$ and relativistic, bosonic $S=1/2$ spinons $z_\alpha$ which carry
the U(1) gauge charge. Upon doping the antiferromagnet in the vicinity of this deconfined critical point with charge carriers of density $x$, we have to include a density $x$ of spinless fermions (`holons'), $\psi_p$, which carry gauge charge ($p=\pm1$) under the same $U(1)$ gauge field. The resulting gauge theory for $z_\alpha$ and $\psi_p$ has a complicated phase diagram as
a function of $g$ and the hole density, which is sketched in Fig.~\ref{phases} and will be partly reviewed in Section~\ref{mod}.
Our attention here will focus on the FL* region of Fig.~\ref{phases}: here the $z_\alpha$ and $\psi_p$ bind to form gauge-neutral fermions of density $x$, which then
form Fermi pockets in the Brillouin-zone near, but not centered at, $(\pm \pi/2, \pm \pi/2)$. The total area enclosed by these pockets is $x$, and hence the Luttinger volume of $1+x$ is not obeyed,
and a FL* phase is realized.

In this paper, we carry out a generalized RPA analysis of a model of this FL* phase interacting via short-range interactions. In most of the parameter space that we have explored, the leading instability in the particle-hole channel is a bond density wave with predominantly $d-$form factor whose wavevector nests the tips of the hole-pockets.\footnote{The density of states, $n(E)=\oint_{E_\k=E}\frac{1}{|\nabla_\k E_\k|}d^2\k$ is high at the tips of the pockets, $E_\k=0$.}
This result is a promising step in the direction of identifying essential characteristics of the normal state which are responsible for giving rise to the BDW instability that is observed experimentally in the underdoped cuprates.

The rest of this paper is arranged as follows: we describe and review a particular route towards constructing an FL* state, starting from a one-band model of electrons coupled to the fluctuations of an AFM order-parameter in sections \ref{mod} and \ref{fl*}. We then setup our computation for determining the charge-ordering instabilities in an FL* interacting via short-range interactions in section \ref{BS}. Finally, we describe our results for the nature of the BDW instabilities, with special emphasis on its wavevector and form factor, in section \ref{res} and conclude with a future outlook in section \ref{dis}. We review some of the previous analysis of density-wave instabilities in metals with large Fermi surfaces interacting via short-range interactions (\cite{SSRP13, AASS14b}) in appendix \ref{FL} in order to highlight the differences with the main results presented in this paper. Appendix \ref{Coulomb} contains a brief discussion of density-wave instabilities in the presence of large Coulomb repulsion.

\section{Model} 
\label{mod}

We begin by summarizing the arguments \cite{RK07,YQSS10} that lead to the phase diagram in Fig.~\ref{phases}.
The starting point is the theory for the electrons, $c_{i\alpha}$ ($\alpha=\uparrow,\downarrow$), hopping on the sites of a square lattice. The electrons are coupled to the fluctuations of an O(3) field, $\n_i$, which describes the local orientation of the antiferromagnetic N$\acute{\tn{e}}$el order at $\K=(\pi,\pi)$. We shall focus on the long-wavelength fluctuations of $\n_i$ while retaining the full lattice dispersions for the fermions. 
The imaginary time Lagrangian, $\L=\L_f+\L_n+\L_{fn}$, is given by,
\beq
\L_f&=&\sum_{i,j}c_{i\alpha}^\dagger\bigg[(\partial_\tau-\mu)\delta_{ij} - t_{ij}\bigg]c_{j\alpha} +\tn{h.c.},\\ 
\L_n&=&\frac{1}{2g}\int d^2\r ~[(\partial_\tau\n)^2+\rm{v}^2(\nabla\n)^2],\\
\L_{fn}&=&\lambda\sum_i e^{i\K\cdot\r_i}~ \n_i\cdot c_{i\alpha}^\dagger \vec\sigma_{\alpha\beta} c_{i\beta}.
\eeq
In the above, $t_{ij}$ represent the hopping matrix elements that give rise to a large Fermi surface, $\mu$ represents the chemical potential, $\lambda$ is an $O(1)$ coupling, $\rm{v}$ represents a characteristic spin-wave velocity and $g$ is used to tune the strength of quantum fluctuations associated with the AFM order parameter. The vector $\n$ satisfies the local constraint $\n_i^2=1$. 

When $g<g_c$, the above model has long-range antiferromagnetic order, $\langle\n\rangle\neq0$ (with a correlation length, $\xi\rightarrow\infty$). In the presence of such long-range order (LRO), the large Fermi surface breaks up into electron and hole pockets. The resulting state in the presence of short-range fermionic interactions could be unstable to other symmetry-broken states; the nature of these instabilities in the particle-hole channel, starting from a reconstructed Fermi surface, have been analyzed \cite{AKB14, ATSS14}. 
As we noted in Section~\ref{intro}, this approach leads to density waves with the $d$-form factor suppressed \cite{AKB14}, consistent with recent observations on the La-based cuprates
\cite{AADH14} which do have magnetic order at low temperatures.

Our interest in the present paper is on the $g \geq g_c$ portion of the phase diagram in Fig.~\ref{phases}, 
which we argue is relevant to the physics of the non-La-based cuprates.
The deconfined critical point $g=g_c$ and $x=0$ is expressed not in terms of the $\n$ fields, but in terms of the spinor $z_\alpha$, with
\beq
\n_i=z_{i\alpha}^*\vec\sigma_{\alpha\beta}z_{i\beta}
\eeq
and a U(1) gauge field $A_\mu$.
At the same time \cite{YQSS10}, one tranforms the underlying electrons, $c_{i\alpha}$, to a new set of spinless Fermions, $\psi_{ip}$,
\beq
c_{i\alpha}&=&R_{\alpha p}^i \psi_{ip},~~~\tn{where}\\
R_{\alpha p}^i&=&\left( \begin{array}{cc}
z_{i\uparrow} & -z_{i\downarrow}^* \\
z_{i\downarrow} & z_{i\uparrow}^*
\end{array} \right), \label{cRz}
\eeq
is a spacetime dependent SU(2) matrix ($\sum_\alpha|z_{i\alpha}|^2=1$) and the fermions $\psi_{ip}$ carry opposite charges $p=\pm 1$ under the same emergent U(1) gauge transformation.  

We begin our discussion of the complex dynamics of $z_\alpha$ and $\psi_p$ by first considering the effect of non-zero $x$ at the critical coupling $g=g_c$.
At very low hole density, each $\psi_p$ fermion can be treated independently of all the others, while interacting with the deconfined gauge theory described by $z_\alpha$
and $A_\mu$. As shown in Ref.~\onlinecite{RK07}, there is an `orthogonality catastrophe' and each $\psi_p$ fermion is effectively localized by
the critical fluctuations. We are not interested here in the very small values of $x$ at which this localization happens, and so will not discuss it further; it is also excluded
from Fig.~\ref{phases} (shown as the grey-shaded region). Moving to higher hole densities, we can assume the holons form a Fermi surface, and this can then quench the $A_\mu$ fluctuations via Landau damping \cite{RKMMSS08,TS08}; the holon Fermi surface is then stable,\footnote{At low temperatures, the holons can pair to form a composite Boson that is neutral under the $A_\mu$ field, condensation of which leads to the holon superconductor \cite{RK08}.} and we obtain the holon metal shown in Fig.~\ref{phases}.

Let us now turn to $g>g_c$. Now there is at least one additional length scale, the spin correlation length, $\xi$. This length should be compared with the spacing between
the holons $\sim 1/\sqrt{x}$. There is actually another significant scale, the length scale at which Landau-damping of the photon sets in; this also diverges as $x \rightarrow 0$, 
and for simplicity we will ignore its difference from the spacing between the holons. When $1/\sqrt{x} \ll \xi$, we revert to the $g=g_c$ situation described in the 
previous paragraph. However, for $\xi \ll 1/\sqrt{x}$, we have to first consider the influence of a non-zero $\xi$ on the gauge theory of the deconfined critical point.
As described in Refs.~\onlinecite{senthil1,senthil2}, here we crossover to a Coulomb phase in which the $A_\mu$ field mediates a logarithmic Coulomb force. 
This Coulomb force binds the $z_\alpha$ and $\psi_p$ quanta into gauge-neutral fermions \cite{RK07,RK08,YQSS10} (an additional attractive force is also provided by the ``Shraiman-Siggia term'' \cite{RK07,YQSS10,SS88}). At longer scales, these gauge-neutral fermions start to notice each
other via the Pauli principle, and so they form Fermi surfaces leading to the FL* state of interest here.\footnote{ The FL* phase considered here is a descendant of the ``holon-hole" metal phase of ref.\cite{RK08}, in the extreme limit where all of the holon states have been depleted into forming gauge-neutral fermions.}

At even longer length scales we have to consider confinement effects from monopoles in the $A_\mu$ gauge field, which are not suppressed in the FL* regime
(but monopoles are suppressed in the holon metal). This we will not do here, leaving it as a difficult but important problem for future study.

\subsection{Fractionalized Fermi liquid (FL*)}
\label{fl*}

As discussed in section \ref{intro} and in Refs.~\onlinecite{YQSS10, EGMSS11}, the emergent photon gives rise to binding of the $\psi_p$ fermions and the $z_\alpha$ spinons into gauge-neutral objects. However, there are two such combinations,
\beq
F_{i\alpha}\sim z_{i\alpha}\psi_{i+},~~G_{i\alpha}\sim \varepsilon_{\alpha\beta}z^*_{i\beta}\psi_{i-},
\eeq
where $\varepsilon_{\alpha\beta}$ is the unit antisymmetric tensor. The physical electronic operator has a non-zero overlap with both of these,
\beq
c_{i\alpha}\equiv Z(F_{i\alpha}+G_{i\alpha}),
\eeq
where $Z$ is a quasiparticle renormalization factor that is nonlocal over $\xi$; this expression differs from the bare relationship
in Eq.~(\ref{cRz}) because we are now dealing with fully renormalized quasiparticles \cite{YQSS10}.
Over distances that are larger than $\xi$, where there is no net AFM order, the $F_\alpha$ and $G_\alpha$ fermions preferentially, but not exclusively, reside on the different sublattice sites.  

Based on symmetry considerations alone, we can write the following effective Hamiltonian for $F_{i\alpha}$ and $G_{i\alpha}$,
\beq
H_{\tn{eff}}=&-&\sum_{i,j} t_{ij}(F_{i\alpha}^\dagger F_{j\alpha} + G_{i\alpha}^\dagger G_{j\alpha}) + \lambda\sum_i e^{i\K\cdot\r_i}(F_{i\alpha}^\dagger F_{i\alpha} - G_{i\alpha}^\dagger G_{i\alpha}) \nonumber\\
&-& \sum_{i,j}\tilde{t}_{ij}(F_{i\alpha}^\dagger G_{j\alpha} + G_{i\alpha}^\dagger F_{j\alpha}).
\eeq
Once again, the $t_{ij}$ represent the hopping matrices corresponding to a large Fermi surface, $\lambda$ represents the potential due to the local AFM order (at distances shorter than $\xi$). The terms proportional to $\tilde{t}_{ij}$ represent the analogs of the ``Shraiman-Siggia" (SS) terms \cite{SS88} which couple the $F$ and $G$ particles. In the absence of these terms, but with $\lambda\neq0$, one obtains hole-like pockets centered at $(\pi/2,\pi/2)$. However, when the SS terms are finite, the pockets can be shifted away from these special points.

It is more convenient now to change basis to a new set of fermionic operators,
\beq
C_{\k\alpha}=\frac{1}{\sqrt{2}}(F_{\k\alpha}+G_{\k\alpha}), ~~D_{\k\alpha}=\frac{1}{\sqrt{2}}(F_{\k+\K\alpha}-G_{\k+\K\alpha}),
\eeq
so that the physical electronic operator $c_{\k\alpha}\simeq (Z/\sqrt{2}) C_{\k\alpha}$. 
The revised Hamiltonian then reads,
\beq
H_\tn{eff}=\sum_\k \bigg[\xi_\k^+ C_{\k\alpha}^\dagger C_{\k\alpha} &+& \xi_{\k+\K}^- D_{\k\alpha}^\dagger D_{\k\alpha}\nonumber\\
-\lambda(C_{\k\alpha}^\dagger D_{\k\alpha} &+& D_{\k\alpha}^\dagger C_{\k\alpha})\bigg],
\label{heff}
\eeq
where the dispersions, $\xi_\k^+,~\xi_\k^-$ are given by,
\beq
\xi_\k^+&=&\ve_\k+\tilde\ve_\k,~~\xi_\k^-=\ve_\k-\tilde\ve_\k,\\
\ve_\k&=&-2t_1(\cos(k_x)+\cos(k_y))-4t_2\cos(k_x)\cos(k_y)\nonumber\\
&&-2t_3(\cos(2k_x)+\cos(2k_y))-\mu,\\
\tilde\ve_\k&=&-\tilde{t}_0-\tilde{t}_1(\cos(k_x)+\cos(k_y)).
\eeq
   
The Green's function for the electronic operator can be obtained from the Hamiltonian in eqn.\ref{heff} and is given by \cite{YQSS10},
\beq
G^c(\k,\omega)&=&\frac{Z^2}{\omega-\xi^+_\k-\lambda^2/[\omega-\xi^-_{\k+\K}]}.
\label{ACL}
\eeq

It is more transparent to rewrite the above Green's function in the following form,
\beq
G^c(\k,\omega)&=&\sum_{\alpha=\pm}\frac{Z_\k^\alpha}{\omega-E^\alpha_\k},~\tn{where}\\
\bigg(\frac{Z^\pm_\k}{Z^2}\bigg)^{-1}&=& 1+\frac{\lambda^2}{(E_\k^\pm-\xi_{\k+\K}^-)^2} ,\\
E_\k^\pm&=&\frac{\xi_\k^+ +\xi_{\k+\K}^-}{2}\pm\sqrt{\bigg(\frac{\xi_\k^+-\xi_{\k+\K}^-}{2}\bigg)^2+\lambda^2}.
\eeq
%The spectral function, $A^c(\k,\omega=0)$, corresponding to the above Green's function is plotted in fig. \ref{akw} for band-structures similar to those of the cuprates. 
In the limit where $\lambda=\tilde{t}_{ij}=0$, one recovers the original large Fermi surface, $\xi_\k$.

\subsection{Charge-order instabilities via T-matrix} 
\label{BS}
Based on the analysis above, we have arrived at a description of the electronic excitations which have been renormalized by the quantum fluctuations of the antiferromagnet. A natural question that we need to address now is whether the resulting state is unstable to other symmetry-broken phases in the presence of short-range interactions. We shall address this question by studying the effect of short-range Coulomb repulsion and AF exchange interaction acting on top of the FL* phase, described by,
\beq
H_C&=&U\sum_i c_{i\uparrow}^\dagger c_{i\uparrow} c_{i\downarrow}^\dagger c_{i\downarrow}  + \sum_{i<j}V_{ij} c_{i\alpha}^\dagger c_{i\alpha} c_{j\beta}^\dagger c_{j\beta},\\
H_J&=&\sum_{i<j}\frac{J_{ij}}{4}\vec\sigma_{\alpha\beta}\cdot\vec\sigma_{\gamma\delta} ~c_{i\alpha}^\dagger c_{i\beta}c_{j\gamma}^\dagger c_{j\delta}.
\label{hjhc}
\eeq

The notation that we shall use to define the interaction parameters from now on is as follows: $J_{ij}\equiv J_{a},~V_{ij}\equiv V_{a}$, where $a (=1,2,3)$ denotes whether $i,~j$ are $1^{st}$, $2^{nd}$, or, $3^{rd}$ nearest neighbors.

Let us now look for the possible charge-ordering instabilities. We will consider the effect of first, second and third neighbor Coulomb and exchange interactions ($i.e.$ $V_\ell,~J_\ell$ with $\ell=1,2,3$). Our generalized order parameter in the particle-hole channel, $P_\Q(\k)$, at a wavevector $\Q$ can be defined as follows (as in eqn.\ref{Dij}):
\beq
\langle c_{i\alpha}^\dagger c_{j\alpha}\rangle = \sum_\Q \bigg[\int_\k P_\Q(\k) e^{i\k\cdot(\r_i-\r_j)} \bigg] e^{i\Q\cdot(\r_i+\r_j)/2}.
\eeq
It is useful to expand $P_\Q(\k)$ in terms of a set of orthonormal basis functions $\phi_\ell(\k)$ as,
\beq
P_\Q(\k)=\sum_\ell \PP_\ell(\Q) \phi_\ell(\k),
\eeq
where we choose a set of $13$ orthonormal basis functions, as described in table \ref{tabbas}.
%%%%%%%%%%%%%%%%%%%%%%%%%%%%%%%%%%%%%%%%%%
\begin{table}[h]
\begin{tabular}{|c||c|c|c||c||c|c|c|}
\hline
$\ell$ & $\phi_\ell(\k)$ & $\overline{J}_\ell$ & $\overline{V}_\ell$ &$\ell$ & $\phi_\ell(\k)$ & $\overline{J}_\ell$ & $\overline{V}_\ell$  \\ \hline\hline
$0$ & 1& 0 & $U$ & & &  & \\ \hline
$1$ & $\cos k_x-\cos k_y$ & $J_1$ & $V_1$ &$7$ & $\sin k_x-\sin k_y$ & $J_1$ & $V_1$    \\ \hline
$2$ & $\cos k_x+\cos k_y$ & $J_1$ & $V_1$ &$8$ & $\sin k_x+\sin k_y$ & $J_1$ & $V_1$   \\ \hline
$3$ & $2\sin k_x\sin k_y$ & $J_2$ & $V_2$ &$9$ & $2\cos k_x\sin k_y$ & $J_2$ & $V_2$    \\ \hline
$4$ & $2\cos k_x\cos k_y$ & $J_2$ & $V_2$ & $10$ & $2\sin k_x\cos k_y$ & $J_2$ & $V_2$   \\ \hline
$5$ & $\cos 2k_x-\cos 2k_y$ & $J_3$ & $V_3$ & $11$ & $\sin 2k_x-\sin 2k_y$ & $J_3$ & $V_3$   \\ \hline
$6$ & $\cos 2k_x+\cos 2k_y$ & $J_3$ & $V_3$ & $12$ & $\sin 2k_x+\sin 2k_y$ & $J_3$ & $V_3$    \\ \hline
\end{tabular}
\caption{Basis functions, $\phi_\ell(\k)$, used for determining the symmetry of the charge-ordering instability. }
\label{tabbas}
\end{table}
%%%%%%%%%%%%%%%%%%%%%%%%%%%%%%%%%%%%%%%%%%
The basis functions from $\ell=0,..,6$ preserve time-reversal symmetry, while the ones from $\ell=7,..,12$ spontaneously break time-reversal symmetry. The remainder of our analysis will be carried out using the T-matrix formalism developed in ref.\cite{AASS14b} to determine the structure of the charge-ordering instability.

The first step in this procedure involves expressing the interaction terms in eqn.\ref{hjhc} as,
\beq
&&H_J+H_C=\sum_{\k,\k',\q}\sum_{\ell=0}^{12}\phi_\ell(\k)\phi_{\ell'}(\k')\bigg[\frac{\overline{J}_\ell}{8}\vec\sigma_{\alpha\beta}\cdot\vec\sigma_{\gamma\delta} c^\dagger_{\k'-\q/2, \alpha}c_{\k-\q/2, \beta}c^\dagger_{\k+\q/2,\gamma}c_{\k'+\q/2, \delta} \nonumber\\
&&~~~~~~~~~~~~~~~~~~~~~~~~~~~~~~~~~~~~~~~~~~+\frac{\overline{V}_\ell}{2} c^\dagger_{\k'-\q/2, \alpha}c_{\k-\q/2, \alpha}c^\dagger_{\k+\q/2,\beta}c_{\k'+\q/2, \beta} \bigg],
\eeq
where $\overline{J}_\ell,~\overline{V}_\ell-$ represent various interaction parameters, as summarised in table \ref{tabbas}.

\begin{figure}
\begin{center}
\includegraphics[width=0.8\columnwidth]{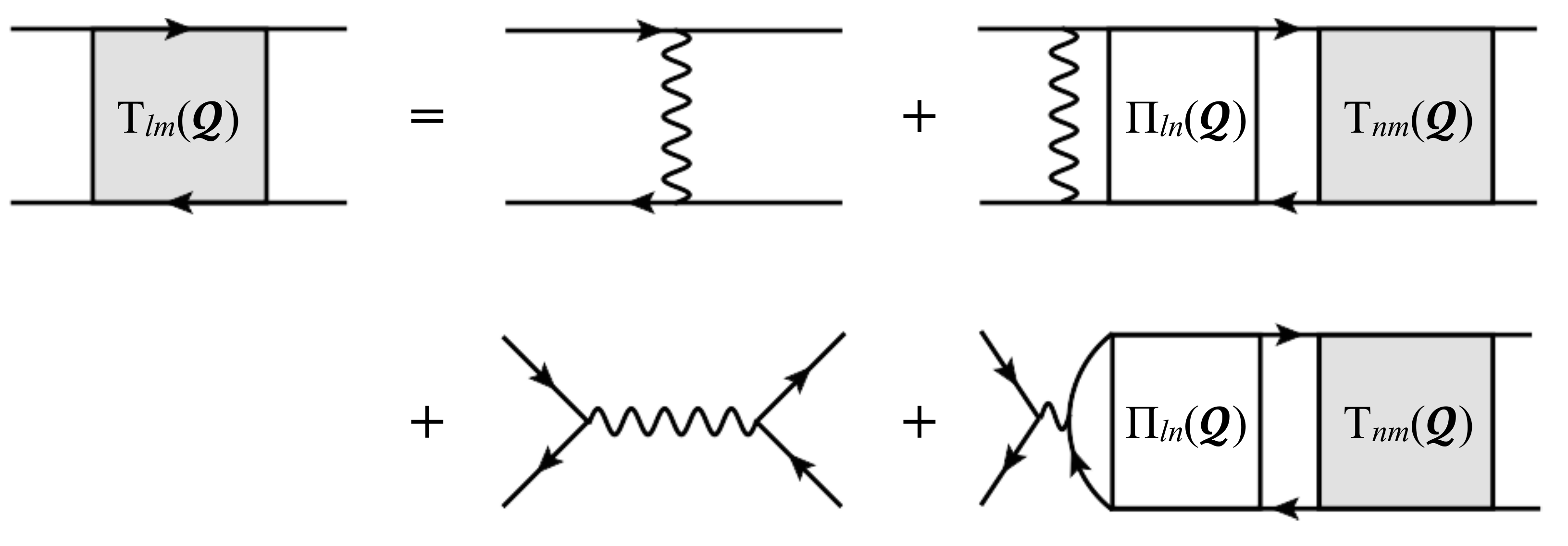}
\end{center}
\caption{The equation for the T-matrix in the particle-hole channel with total momentum $\Q$.}
\label{CO}
\end{figure}

After summing all the ladder diagrams in the particle-hole channel, we obtain (see fig.\ref{CO}),
\beq
T_{\ell m}(\Q)&=&\bigg(\frac{3}{4}\overline{J}_\ell + \overline{V}_\ell \bigg)\delta_{\ell m}-2\delta_{\ell,0}\delta_{m,0}W(\Q)\nonumber\\
&+&\frac{1}{2}\sum_{n=0}^{12}\bigg(\frac{3}{4}\overline{J}_\ell + \overline{V}_\ell \bigg) \Pi_{\ell n}(\Q) T_{nm}(\Q)- \delta_{\ell,0}\sum_{n=0}^{12}W(\Q) \Pi_{0,n}(\Q) T_{nm}(\Q),
\label{Tlm}
\eeq 
where,
\beq
W(\Q)\equiv\sum_{\ell=0}^{12}\overline{V}_\ell \phi_\ell(0)\phi_\ell(\Q)
\eeq
arises from the direct interaction, and the polarizabilities are given by,
\beq
\Pi_{\ell m}(\Q)&=&2\sum_{\k}\phi_\ell(\k)\phi_m(\k) \Pi_\Q(\k),\\
\Pi_\Q(\k) &=& \sum_{\alpha,\beta=\pm}Z^\alpha_{\k+\Q/2} Z^\beta_{\k-\Q/2} \frac{f(E_{\k+\Q/2}^\alpha)-f(E_{\k-\Q/2}^\beta)}{E_{\k-\Q/2}^\beta-E_{\k+\Q/2}^\alpha}.
\eeq
In the above, $f(...)$ represents the Fermi-Dirac distribution function.

%%%%%%%%%%%%%%%%%%%%%%%%%%%%%%%%%%%%%%%%%%%%%%%%%%%%%%%%%%%
%%%%%%%%%%%%%%%%%%%%%%%%%%%%%%%%%%%%%%%%%%%%%%%%%%%%%%%%%%%

From eqn.\ref{Tlm}, it is straightforward to see that the charge-ordering instability is determined by the lowest eigenvalues, $\lambda_\Q$, of the matrix $\M_{\ell,\ell'}(\Q)$,
\beq
\M_{\ell,\ell'}(\Q)=\delta_{\ell\ell'}-\frac{1}{2}\bigg(\frac{3}{4}\overline{J}_\ell + \overline{V}_{\ell} \bigg)\Pi_{\ell\ell'}(\Q) + \delta_{\ell, 0}W(\Q)\Pi_{0\ell'}(\Q),
\eeq
and the $\PP_{\ell'}(\Q)$ are determined by the corresponding right eigenvector.

In the remainder of the paper, we shall investigate the nature of these instabilities by studying the eigenvalues and eigenvectors corresponding to the matrix $\M(\Q)$ as a function of $\Q$. 

\section{Results}
\label{res}
\begin{figure}[ht!]
\begin{center}
\includegraphics[scale = 0.5]{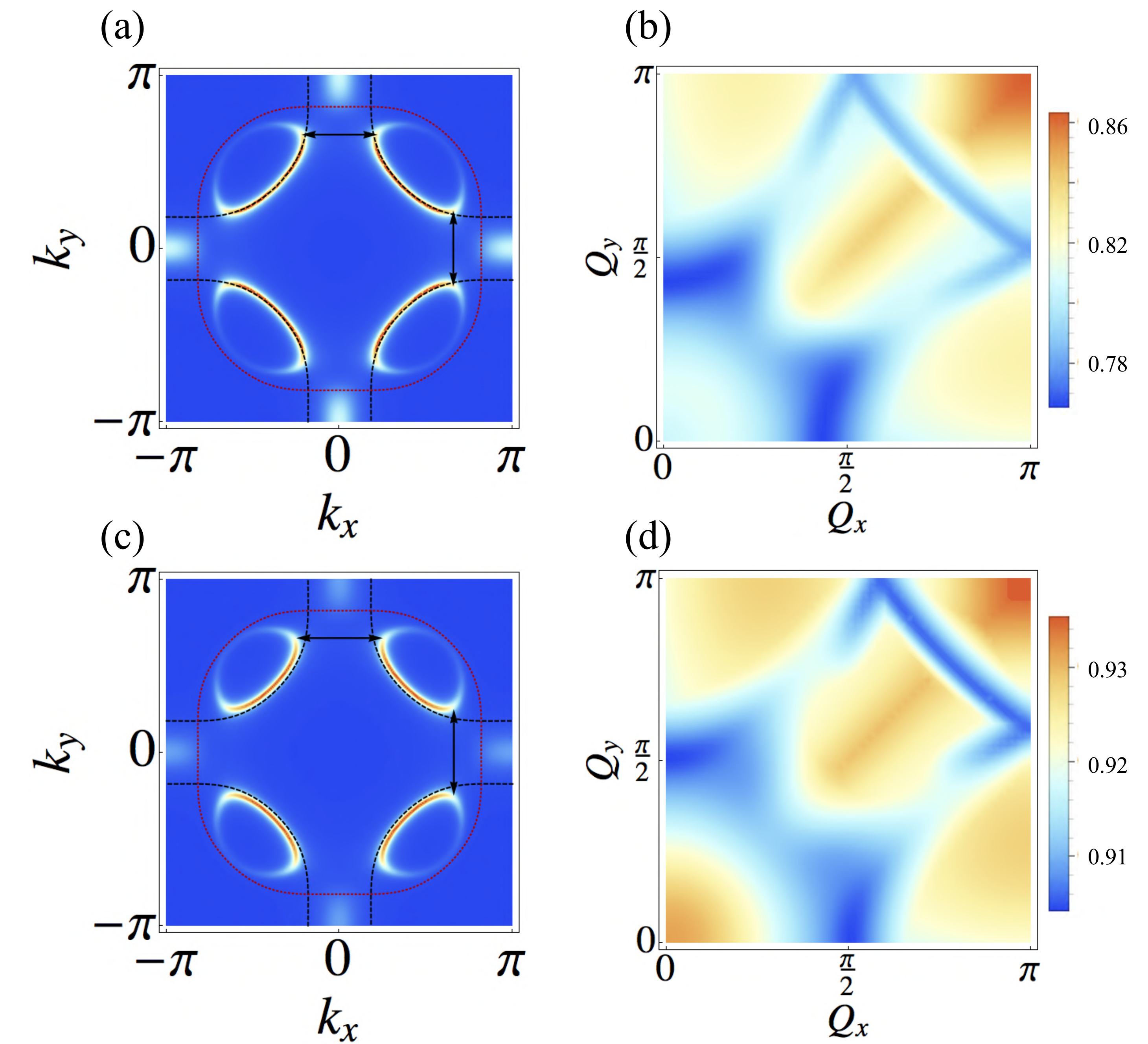}
\end{center}
\caption{The spectral function, $A^c(\k,\omega=0)$, for bare hopping parameters $t_1=1.0,~t_2=-0.32,~ t_3 = 0.128,~\mu= -1.11856$ and FL* parameters: (a) $\tilde{t}_0=-0.5t_1,~\tilde{t}_1=0.4t_1,~ \lambda=0.6t_1$, and, (c) $\tilde{t}_0=-0.5t_1,~\tilde{t}_1=0.6t_1,~ \lambda=0.75t_1$. The black dashed lines represent, $\e_\k=0$, and the red dotted lines represent, $\e_{\k+\K}=0$. The lowest eigenvalue, $\lambda_\Q$ as a function of $\Q$ at a temperature $T=0.06$ are shown in (b), (d) for the FL* states in (a), (c) respectively. The wavevectors corresponding to the minimum eigenvalues are shown as the black arrows. The exchange interaction parameters are given by (b) $J_1=1.0,~J_2=J_3=0.05$, and, (d) $J_1=0.5,~J_2=J_3=0.05$. $U=V_1=V_2=V_3=0$ for both cases. We have put in a finite imaginary part ($=0.1t_1$) in the Green's function for visualization purpose. }
\label{lambQ_J}
\end{figure}

\begin{figure}[ht!]
\begin{center}
\includegraphics[scale = 0.5]{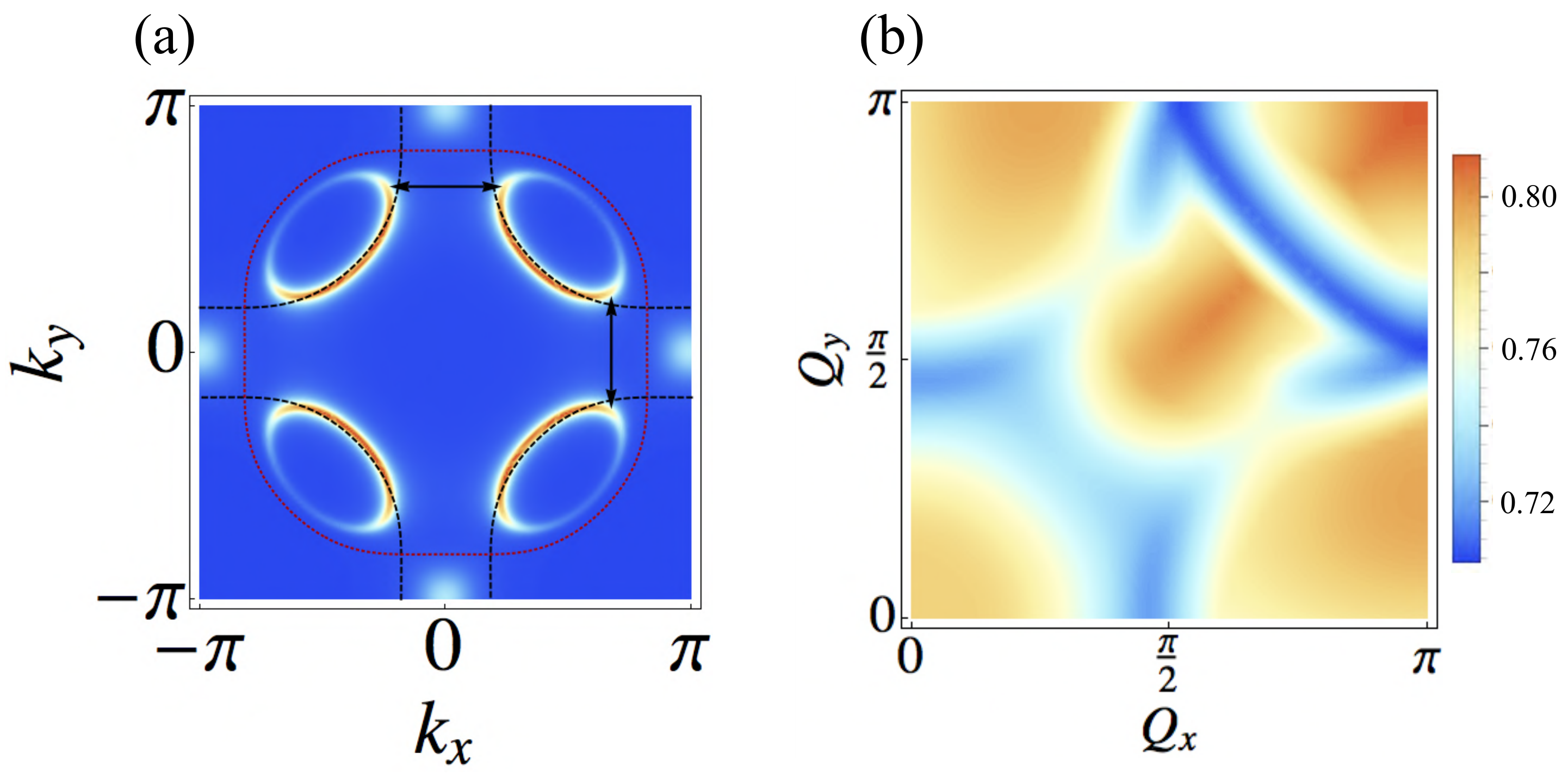}
\end{center}
\caption{(a) The spectral function, $A^c(\k,\omega=0)$, for bare hopping parameters $t_1=1.0,~t_2=-0.32,~ t_3 = 0.128,~\mu= -1.11856$ and FL* parameters: $\tilde{t}_0=-0.2t_1,~\tilde{t}_1=-0.1t_1,~ \lambda=0.75t_1$. The black dashed lines represent, $\e_\k=0$, and the red dotted lines represent, $\e_{\k+\K}=0$. The lowest eigenvalue, $\lambda_\Q$ as a function of $\Q$ at a temperature $T=0.06$ is shown in (b). The wavevectors corresponding to the minimum eigenvalue are shown as the black arrows. The interaction parameters are given by $J_1=1.0,~J_2=0.1,~J_3=0.05$, and, $U=0,~V_1=0.05,~V_2=V_3=0.01$. We have put in a finite imaginary part ($=0.1t_1$) in the Green's function for visualization purpose.}
\label{lambQ_JV}
\end{figure}

We have analyzed the lowest eigenvalues, $\lambda_\Q$, as a function of $\Q$ for a variety of interaction and FL* parameters (see figs. \ref{lambQ_J}, \ref{lambQ_JV}). These results should be contrasted with the eigenvalues obtained for instabilities of metals with a large Fermi surface, interacting via short-ranged interactions, as shown in appendix \ref{FL} and earlier works \cite{SSRP13,AASS14b}. 

Let us now start by exploring the nature of these instabilities in the presence of purely exchange interactions, $i.e.$ set $U=V_1=V_2=V_3=0$. We plot $\lambda_\Q$ as a function of $\Q$ in fig.\ref{lambQ_J} for two different choice of FL* and exchange-interaction parameters, $\{J_1,~J_2,~J_3\}$. For the FL* state shown in fig.\ref{lambQ_J}(a), the charge-ordering eigenvalues are displayed in fig.\ref{lambQ_J}(b). Note that the global-minimum at $\Q=(\pm Q_0,\pm Q_0)$, which is a robust feature of an instability arising out of a large FS (with $\tilde{t}_0=\tilde{t}_1=\lambda=0$), has disappeared (see fig. \ref{FLlambQ_J}(a) in appendix \ref{FL}; Refs. \cite{SSRP13,AASS14b}). Instead, there are now ridges of instability that extend starting from wavevectors of the type $(Q_0,0)$ and $(0,Q_0)$. Interestingly, the lowest eigenvalue is shifted slightly away from the axis and corresponds to $\Q^*=(\pm 11\pi/25 ,\pm 3\pi/50),~(\pm 3\pi/50,\pm 11\pi/25)$. However, the eigenvalue, $\lambda_{\Q^{**}}$, for the state on the axis with $\Q^{**}=(\pm 11\pi/25 ,0),~(0,\pm 11\pi/25)$, is infinitesimally close to the lowest eigenvalue.\footnote{where the difference, $|\lambda_{\Q^{**}}-\lambda_{\Q^*}|\simeq 10^{-4}$.} The charge-ordering eigenvectors for these states are given by,
\beq
P_{\Q^*}(\k)=&-&0.996[\cos k_x-\cos k_y] + 0.079 [\cos k_x+\cos k_y] \nonumber\\
&+& 0.017  [2\cos k_x\cos k_y]\nonumber\\
 &-& 0.027 [\cos 2k_x - \cos 2k_y] - 0.014  [\cos 2k_x + \cos 2k_y], \\
P_{\Q^{**}}(\k)=&+&0.996[\cos k_x-\cos k_y] - 0.084 [\cos k_x+\cos k_y] \nonumber\\
&-& 0.017  [2\cos k_x\cos k_y]\nonumber\\
 &+& 0.027 [\cos 2k_x - \cos 2k_y] + 0.014  [\cos 2k_x + \cos 2k_y], 
\eeq 
both of which predominantly have a $d-$wave component (and a tiny $s'-$ component). 

While the eigenvalue analysis doesn't directly tell us how the wavevector of the leading instability relates to the underlying Fermi surface geometry, it is reasonable to associate it with points in the Brillouin-zone which have a high joint density of states. It is then straightforward to see that $\Q^{**}$ connects the tips of the pockets (shown as the b arrows in fig.\ref{lambQ_J}a). There is also a secondary ridge of instability, which corresponds to a set of local but not global minima, extending from approximately $\Q\approx(\pi/2,\pi)$ to $\Q\approx(\pi,\pi/2)$. Upon close inspection, we realize that such a ridge exists even for the large Fermi surface computation (fig.\ref{FLlambQ_J}a), though the eigenvalues there were significantly larger than the one corresponding to the global minimum. These BDW states contain an admixture of $d-$ and $s'-$ form factors. However, the states marked in yellow in the vicinity of $\Q=(\pi,\pi)$ and $\Q=(\pi,0),~(0,\pi)$ break time-reversal symmetry and correspond to states that have spontaneous currents. 
%%%%%%%%%%%%%%%%%%%%%%%%%%%%%%%%%%%%%%%%%%%%
%%%%%%%%%%%%%%%%%%%%%%%%%%%%%%%%%%%%%%%%%%%%%

We can repeat a similar analysis for other FL* and interaction parameters, as shown in fig.\ref{lambQ_J}(c), (d). For the FL* state shown in fig.\ref{lambQ_J}(c), the lowest eigenvalue corresponds to $\Q^*=(\pm\pi/2,0),~(0,\pm\pi/2)$. Though the particular wavevector in this case corresponds to a period-4 CDW, this is entirely a coincidence; $\Q^*$ happens to connect the tips of the pockets (shown as black arrows in fig.\ref{lambQ_J}c). The charge-ordering eigenvector for this state is given by,
\beq
P_{\Q^*}(\k)=&+&0.993 [\cos k_x-\cos k_y] - 0.095 [\cos k_x+\cos k_y] \nonumber\\
&-& 0.032 [2\cos k_x\cos k_y]\nonumber\\
 &+& 0.050 [\cos 2k_x - \cos 2k_y] + 0.027  [\cos 2k_x + \cos 2k_y],
\eeq 
which, once again, predominantly has a $d-$wave component. The secondary ridge of instability appears here as well, extending from approximately $\Q\approx(29\pi/50,\pi)$ to $\Q\approx(\pi,29\pi/50)$, and contains an admixture of $d-$ and $s'-$ form factors. The states marked in yellow in the vicinity of $\Q=(\pi,\pi)$ and $\Q=(\pi,0),~(0,\pi)$ continue to  break time-reversal symmetry. 

The results for both of these cases above are very similar, with one major qualitative difference. In the first case, the region around $\Q=0$ has a local ``valley" of instability, where the state at $\Q=0$ corresponds to a nematic instability with a purely $d-$form factor. This is no longer true for the second case considered here.   
%%%%%%%%%%%%%%%%%%%%%%%%%%%%%%%%%%%%%%%%%%%%
%%%%%%%%%%%%%%%%%%%%%%%%%%%%%%%%%%%%%%%%%%%%%

Let us now study the problem in the presence of Coulomb repulsion --- the results are displayed in fig.\ref{lambQ_JV}. Interestingly, in this case for the particular choice of parameters, the leading instability in the particle-hole channel for the FL* state in fig.\ref{lambQ_JV}(a) corresponds to $\Q^*= (\pm13\pi/25,\pm\pi)$, $(\pm\pi,\pm13\pi/25)$. The charge-ordering eigenvector for this state is given by,
\beq
P_{\Q^*}(\k)=&-& 0.184 - 0.694[\cos k_x-\cos k_y] -  0.694[\cos k_x+\cos k_y] \nonumber\\
&+& 0.021 [2\sin k_x\sin k_y]\nonumber\\
 &-& 0.006 [\cos 2k_x - \cos 2k_y] - 0.033   [\cos 2k_x + \cos 2k_y].
\eeq 
However, as can be seen from fig.\ref{lambQ_JV}(b), there is a subleading instability to a BDW with  $\Q^{**}=(\pm23\pi/50,0)$, $(0,\pm23\pi/50)$, whose eigenvalue, $\lambda_{\Q^{**}}$, is in fact very close to $\lambda_{\Q^*}$. The corresponding eigenvector is given by,
\beq
P_{\Q^{**}}(\k)=&-& 0.012 - 0.986[\cos k_x-\cos k_y] +  0.162[\cos k_x+\cos k_y] \nonumber\\
&+& 0.024 [2\cos k_x\cos k_y]\nonumber\\
 &-& 0.032 [\cos 2k_x - \cos 2k_y] - 0.012   [\cos 2k_x + \cos 2k_y],
\eeq
which, not surprisingly, has a predominantly $d-$form factor. 

The remaining features in the $\lambda_\Q - \Q$ phase diagram remain identical to the earlier results shown in fig.\ref{lambQ_J}(b). 

We investigate the nature of the instabilities for two more cases in appendix \ref{Coulomb}, when $U=2J_1\gg V_1$ and $U\sim J_1\sim V_1$. The nature of the leading instabilities in these two cases is different; in the first case it is a ``staggered-flux" state (though away from $(\pi,\pi)$) that breaks time reversal symmetry, while in the second case it leads to a conventional charge-density wave at $(\pi,\pi)$.
However, within the region of small $|\Q|$ of interest to us here, there remains a local instability to 
a $d$-form factor density wave with wavevectors of the form $(0, Q_0)$ and $(Q_0, 0)$, similar to those found above.

\section{Discussion}
\label{dis}
We have argued here that the incommensurate charge density wave state in the underdoped cuprates acts as a window into the exotic ``normal" phase out of which it emerges. The traditional weak-coupling computations that start with a large Fermi surface give rise to a density wave instability with diagonal wavevectors of the form $(\pm Q_0,\pm Q_0)$, which is in disagreement with the experimental observations. A promising candidate for the normal state of the underdoped cuprates is a U(1)-FL*, where the electrons are coupled to the fractionalized excitations of a quantum fluctuating antiferromagnet. In this paper, we have investigated the charge-ordering instabilities of such a FL* state, and have identified the leading instability of the FL* state in the particle-hole channel. In most of the cases considered here, this leads to a $d-$form factor bond density wave with wavevectors of the form $\Q=(\pm Q_0,0),~(0,\pm Q_0)$, where $Q_0$ is related to a geometric property of the Fermi surfaces of the FL* state. Moreover, as a function of increasing doping, as the size of the hole-pockets increase, the magnitude of the wavevector that nests the tips of the pockets decreases. This agrees with the trend that has been seen in experiments, where the BDW wavector is a decreasing function of the increasing hole-doping \cite{blackburn13}.

While the identification of the correct BDW starting from a more exotic normal state is an interesting result, there are other implications at low temperatures of having a parent FL*. The vanilla FL* state has hole pockets with non-zero (but small) photoemission intensity on the `back side' {\it i.e.\/} outside the first
antiferromagnetic Brillouin zone, and the photoemission observations of Ref.~\onlinecite{Yang11} have been argued to be consistent with this.
However, the superconducting state descending from such a FL* metal has 8 nodal points \cite{EGMSS11}, and no signs of such a feature
have been experimentally detected. But, it should be noted that U(1)-FL* is
ultimately unstable to confinement, because in the absence of fermions carrying the U(1) gauge charge, monopoles must
proliferate at large enough scales. It is expected that this crossover to confinement will resolve the experimental conflicts of the
FL* state. Moreover, with the presence of an incommensurate BDW, there are no issues with the conventional Luttinger
theorem, and so the crossover to confinement can happen without the need for further symmetry breaking.

Finally, we note implications for experiments on the cuprates in high magnetic fields. In the light of the above results, it would be interesting to investigate how the BDW derived from an FL* evolves as a function of magnetic field, and in particular, 
determine its relation to the observed quantum oscillations at intermediate and high fields \cite{LT07a, LT07b, LT09, NHSS11, NHSS12, MG13, SS14,LT14}. Computations starting from a large Fermi surface in the presence of long-range charge order show a nodal electron-pocket \cite{NHSS11, NHSS12,SS14, LT14,AADCSS} and the more recently discovered hole-pockets \cite{LT14,AADCSS}; it is not implausible that similar results will also be obtained starting from U(1)-FL*, especially after the crossover to confinement has
been accounted for. Moreover, in the FL* framework the full large Fermi surface is never recovered, because even in the absence of 
charge order the spin liquid suppresses much of the Fermi surface: this may be a way of reconciling with
 thermodynamic measurements of the specific-heat \cite{GB} and the spin-susceptibility \cite{KS}. 

\acknowledgements
We thank A.~Allais, W.~A.~Atkinson, A.~Chubukov, J.~C.~Seamus Davis, D.~G.~Hawthorn, J.~E.~Hoffman, P.~Johnson, Jia-Wei Mei, M.~Randeria, T.~Senthil, B.~Swingle and A.~Thomson for interesting 
discussions. D.C. is supported by the Harvard-GSAS Merit Fellowship, and acknowledges the ``Boulder summer school for condensed matter physics - Modern aspects of Superconductivity", where some preliminary ideas for this work were formulated. 
This research was 
supported by the NSF under Grant DMR-1360789, the Templeton foundation, and MURI grant W911NF-14-1-0003 from ARO.
It was also supported by the Perimeter Institute for Theoretical Physics; research at Perimeter Institute is supported by the Government of Canada through Industry Canada and by the Province of Ontario through the Ministry of Research and Innovation.

\appendix
\section{CDW instabilities of FL}
\label{FL}

We plot the lowest eigenvalues, $\lambda_\Q$, for the charge-ordering instabilities of the large FS corresponding to parameters of figs.\ref{lambQ_J}, \ref{lambQ_JV} in fig.\ref{FLlambQ_J} below.
\begin{figure}[ht!]
\begin{center}
\includegraphics[scale = 0.45]{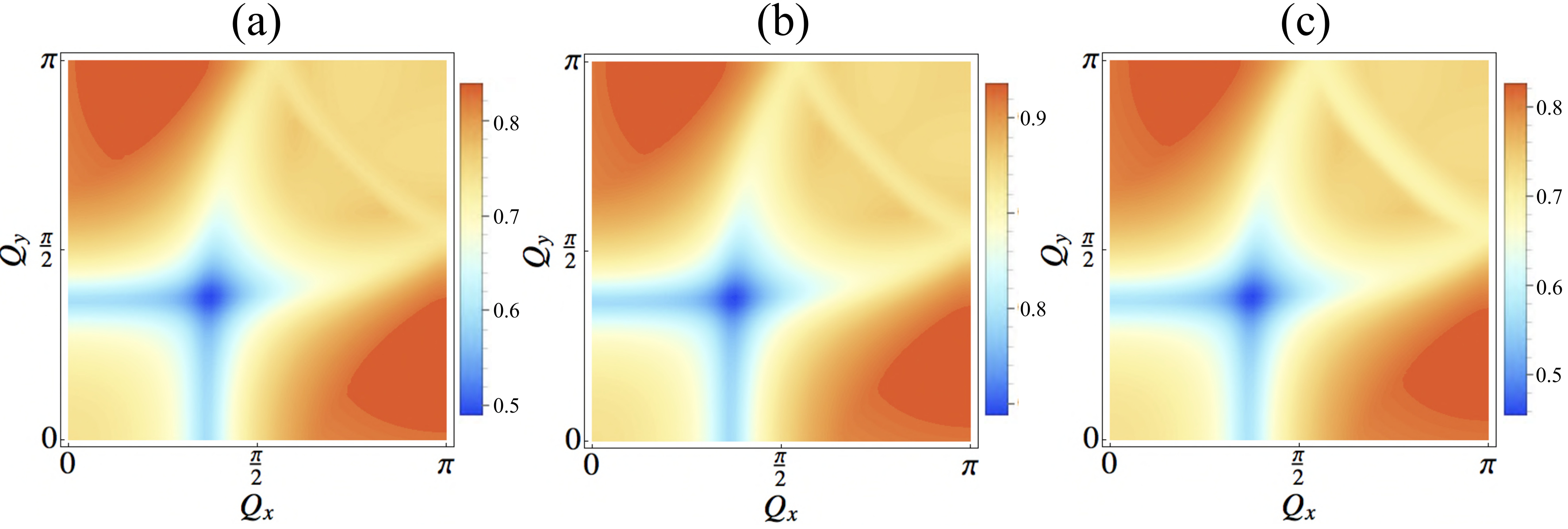}
\end{center}
\caption{The lowest eigenvalues, $\lambda_\Q$, as a function of $\Q$ at a temperature $T=0.06$ for the large FS corresponding to $t_1=1.0,~t_2=-0.32,~ t_3 = 0.128,~\mu= -1.11856$. These computations have been carried out as a limit of the FL* computation with $\tilde{t}_0=0,~\tilde{t}_1=0,~ \lambda=0$. The state with the diagonal wavevector is the leading instability in all the cases. The interaction parameters are given by: (a) $J_1=1.0,~J_2=J_3=0.05, U=V_1=V_2=V_3=0$, (b) $J_1=0.5,~J_2=J_3=0.05,~U=V_1=V_2=V_3=0$, and, (c) $J_1=1.0,~J_2=0.1,~J_3=0.05,~U=0,~V_1=0.05,~V_2=V_3=0.01$. }
\label{FLlambQ_J}
\end{figure}
For the parameters in fig.\ref{FLlambQ_J}(a), the minimum eigenvalue occurs at $\Q^*=(\pm 19\pi/50,\pm19\pi/50)$ and the corresponding eigenvector is given by,
\beq
P_{\Q^*}=-0.999 [\cos k_x-\cos k_y] - 0.014 [\cos(2k_x)-\cos(2k_y)].
\eeq
For the parameters in figs.\ref{FLlambQ_J}(b), (c), the minimum eigenvalue occurs at the same value of $\Q^*$ as above and the form of the eigenvector remains almost identical. This isn't surprising, since $\Q^*$ is determined by the separation between the ``hot-spots".

\section{Instabilities in the presence of strong Coulomb repulsion}
\label{Coulomb}
\begin{figure}[ht!]
\begin{center}
\includegraphics[scale = 0.45]{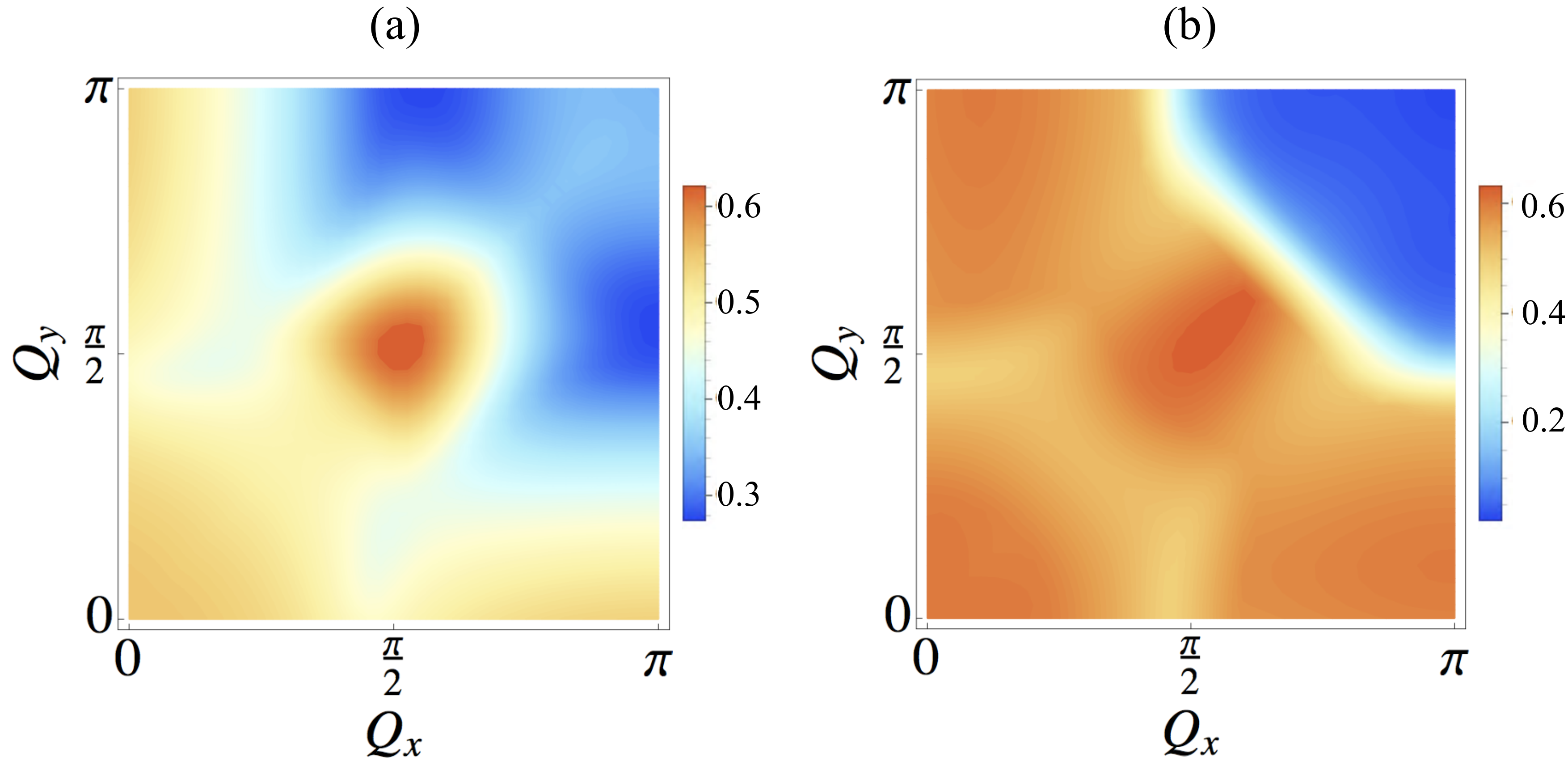}
\end{center}
\caption{The lowest eigenvalues, $\lambda_\Q$, as a function of $\Q$ at a temperature $T=0.06$ for the FL* state shown in fig.\ref{lambQ_JV}(a), corresponding to $t_1=1.0,~t_2=-0.32,~ t_3 = 0.128,~\mu= -1.11856,~\tilde{t}_0=-0.2t_1,~\tilde{t}_1=-0.1t_1,~ \lambda=0.75t_1$. The interaction parameters are given by: (a) $J_1=1.0,~J_2=0.1,~J_3=0.05,~U=2.0,~V_1=0.1,~V_2=V_3=0.01$, (b) $J_1=1.0,~J_2=0.1,~J_3=0.05,~U=1.0,~V_1=0.7,~V_2=V_3=0.01$. }
\label{coul}
\end{figure}

In this appendix, we present some additional results for the weak-coupling density-wave instabilities in the presence of strong Coulomb repulsion ($U\sim J_1$). The leading instabilities that we obtain are not bond-density waves of the type discussed in the main text. We explore the instabilities for the FL* state whose spectral weight is shown in fig.\ref{lambQ_JV}(a) for different combinations of the Coulomb repulsion parameters. The lowest eigenvalues, $\lambda_\Q$ as a function of $\Q$ are shown in fig.\ref{coul}. 

We start with the case when $U=2J_1\gg V_1$ (fig.\ref{coul}a). We find that the leading instability corresponds to $\Q^*=(\pm11\pi/20,\pm\pi),~(\pm\pi,\pm11\pi/20)$ and the eigenvector for this state is predominantly given by $P_\Q(\k)=\sin k_x - \sin k_y$. This state therefore breaks time-reversal symmetry and is a generalized version of the ``staggered-flux" state, but at a wavevector away from $(\pi,\pi)$.

We next consider the case when $U=J_1\sim V_1$ (fig.\ref{coul}b). In this case, we find that the leading instability occurs at $\Q^*=(\pi,\pi)$ and the eigenvector for this state is given by $P_\Q(\k)=1$. This state is therefore just a conventional $(\pi,\pi)$ charge-density wave, that we should naively expect to arise due a large $V_1$.

However, we note that if we focus on the region of the Brillouin zone with smaller $|\Q|$, the $d$-form factor density
waves found in the body of the paper with smaller Coulomb repulsion are present also in the cases considered in this appendix.
Thus the Coulomb repulsion has little direct effect on the $d$-form factor density waves, but, in the present simple RPA framework, 
it can induce other instabilities which are not of current experimental interest.


\begin{thebibliography}{99}
\bibitem{Ghi12} G. Ghiringelli {\it et al.\/}, Science {\bf 337}, 821 (2012).
\bibitem{DGH12} A.J. Achkar {\it et al.\/}, Phys. Rev. Lett. {\bf 109}, 167001 (2012).
\bibitem{SH12} J. Chang {\it et al.\/}, Nat. Phys. {\bf 8}, 534 (2012).
\bibitem{comin13} R. Comin {\it et al.\/}, Science {\bf 343}, 390 (2014).
\bibitem{neto13} E.H. da Silva Neto {\it et al.\/}, Science {\bf 343}, 393 (2014).
\bibitem{DGH13} A.J. Achkar, F. He, R. Sutarto, J. Geck, H. Zhang, Y.-J. Kim, and D. G. Hawthorn, Phys. Rev. Lett. {\bf 110}, 017001 (2013).
\bibitem{MHJ11} T. Wu, H. Mayaffre, S. Kramer, M. Horvatic, C. Berthier, W. N. Hardy, R.Liang, D. A. Bonn and M.
H. Julien, Nature {\bf 477}, 191 (2011).
\bibitem{MHJ13} T. Wu, H. Mayaffre, S. Kramer, M. Horvatic, C. Berthier, P.L. Kuhns, A.P. Reyes, R. Liang, W.N.
Hardy, D.A. Bonn and M.-H. Julien, Nat. Comms. {\bf 4}, 2113 (2013). 
\bibitem{CP13} D. LeBeouf, S. Kramer, W. N. Hardy, R. Liang, D. A. Bonn and Cyril Proust, Nature Phys. {\bf 9}, 79 (2013).
\bibitem{JH02} J. E. Hoffman, E. W. Hudson, K. M. Lang, V. Madhavan, H. Eisaki, S. Uchida and J. C. Davis, Science {\bf 295}, 466 (2002).
\bibitem{AY04} M. Vershinin, S. Misra, S. Ono, Y. Abe, Yoichi Ando, and A.~Yazdani, Science {\bf 303}, 1995 (2004).
\bibitem{JSD1011} M. J. Lawler, K. Fujita, J. Lee, A. R. Schmidt, Y. Kohsaka, C. K. Kim, H. Eisaki, S. Uchida, J. C.
Davis, J. P. Sethna and Eun-Ah Kim, Nature {\bf 466}, 347 (2010); A. Mesaros, Mesaros, K. Fujita, H. Eisaki, S.
Uchida, J. C. Davis, S. Sachdev, J. Zaanen, M. J. Lawler and Eun-Ah Kim, Science {\bf 333}, 426 (2011).
\bibitem{SSJSD14} K.~Fujita, M.~H.~Hamidian {\it et al.\/}, PNAS {\bf 111}, E3026 (2014).
\bibitem{comin2} R.~Comin {\it et al.\/}, arXiv:1402.5415.
\bibitem{MMSS10} M.~A.~Metlitski and S. Sachdev, Phys. Rev. B {\bf 82}, 075128 (2010).
\bibitem{metzner} T.~Holder and W.~Metzner, Phys. Rev. B {\bf 85}, 165130 (2012); 
C.~Husemann and W.~Metzner, Phys. Rev. B {\bf 86}, 085113 (2012).
\bibitem{yamase} M.~Bejas, A.~Greco, and H.~Yamase, Phys. 
Rev. B {\bf 86}, 224509 (2012).
\bibitem{SSRP13} S. Sachdev and R. LaPlaca, Phys. Rev. Lett. {\bf 111}, 027202 (2013).
\bibitem{kee} Hae-Young Kee, C.~M.~Puetter, and D.~Stroud, J. Phys.: Condens. Matter {\bf 25}, 202201 (2013).
\bibitem{SSJS14} J.D. Sau and S. Sachdev, Phys. Rev. B {\bf 89}, 075129 (2014).
\bibitem{DHL13} J.C. Davis and D.H. Lee, Proc. Natl. Acad. Sci. 110, 17623 (2013).
\bibitem{HMKE} K.B. Efetov, H. Meier and C. Pepin, Nat. Phys. {\bf 9}, 442 (2013).
\bibitem{DCSS14} D. Chowdhury and S. Sachdev, Phys. Rev. B {\bf 90}, 134516 (2014). 
\bibitem{HMKE13} H. Meier, C. Pepin, M. Einenkel and K.B. Efetov, Phys. Rev. B {\bf 89}, 195115 (2014).
\bibitem{HF14} V. S. de Carvalho and H. Freire, arXiv:1402.4820.
\bibitem{AASS14} A. Allais, J. Bauer and S. Sachdev, Phys. Rev. B {\bf 90}, 155114 (2014).
\bibitem{AASS14b} A. Allais, J. Bauer and S. Sachdev, Ind. J. Phys. {\bf 88}, 905 (2014).
\bibitem{YWAC14} Y. Wang and A. Chubukov, Phys. Rev. B {\bf 90}, 035149 (2014).
\bibitem{norman14} A. Melikyan and M. R. Norman, Phys. Rev. B {\bf 89}, 024507 (2014).
\bibitem{ATAC14} A.~M~Tsvelik and A.~V.~Chubukov, Phys. Rev. B {\bf 89}, 184515 (2014).
\bibitem{bulut} S.~Bulut, W.~A.~Atkinson, and A.~P.~Kampf, Phys. Rev. B {\bf 88}, 155132 (2013).
\bibitem{AKB14} W.~A.~Atkinson, A.~P.~Kampf, and S.~Bulut, arXiv:1404.1335.
\bibitem{EAK14} M.~H.~Fischer, Si Wu, M.~Lawler, A.~Paramekanti, and Eun-Ah Kim, arXiv:1406.2711
\bibitem{AADH14} A.~J.~Achkar, F. He, R. Sutarto, C. McMahon, M. Zwiebler, M. Hucker, G. D. Gu, R. Liang, D. A. Bonn, W. N. Hardy, J. Geck, D. G. Hawthorn, arXiv:1409.6787.
\bibitem{ATSS14} A.Thomson and S. Sachdev, arXiv:1410.3483.
\bibitem{SSMMMP12} S. Sachdev, M.~A.~Metlitski and M. Punk, Journal of Physics: Condensed Matter {\bf 24}, 294205 (2012).
\bibitem{TSSSMV} T. Senthil, S. Sachdev and M. Vojta , Phys. Rev. Lett. {\bf 90}, 216403 (2003); T. Senthil, M. Vojta and S. Sachdev, Phys. Rev. B {\bf 69}, 035111 (2004).
\bibitem{LZJM14} Long Zhang and Jia-Wei Mei, arXiv:1408.6592.
\bibitem{YRZ} Kai-Yu Yang, T. M. Rice and Fu-Chun Zhang, 
%Phenomenological theory of the pseudogap state, 
Phys. Rev. B {\bf 73}, 174501 (2006).
\bibitem{YQSS10} Y. Qi and S. Sachdev, Phys. Rev. B {\bf 81}, 115129 (2010).
\bibitem{Wen12} Jia-Wei Mei, Shinji Kawasaki, Guo-Qing Zheng, Zheng-Yu Weng, and Xiao-Gang Wen, 
%Luttinger-volume violating Fermi liquid in the pseudogap phase of the cuprate superconductors, 
Phys. Rev. B {\bf 85}, 134519 (2012).
\bibitem{MV12} M. Vojta, Physica C {\bf 481}, 178 (2012); M. Vojta and O. Rosch, Phys. Rev. B {\bf 77}, 094504 (2008).
\bibitem{RK07} R.~K.~Kaul, A.~Kolezhuk, M.~Levin, S.~Sachdev, and T.~Senthil, Phys. Rev. B {\bf 75}, 235122 (2007).
\bibitem{RK08} R.~K.~Kaul, Y.-B.~Kim, S.~Sachdev, and T.~Senthil, Nat. Phys. {\bf 4}, 28 (2008).
\bibitem{SS09} S.~Sachdev,  M.~A.~Metlitski, Y.~Qi, and C.~Xu, Phys. Rev. B {\bf 80}, 155129 (2009).
\bibitem{DCSS14c} D.~Chowdhury and S.~Sachdev, arXiv:1412.1086.
\bibitem{EGMSS11} E.G. Moon and S. Sachdev, Phys. Rev. B {\bf 83}, 224508 (2011).
\bibitem{MPSS12} M. Punk and S. Sachdev, Phys. Rev. B {\bf 85}, 195123 (2012).
\bibitem{senthil1}  T.~Senthil, A.~Vishwanath, L.~Balents, S.~Sachdev, and
M.~P.~A.~Fisher, Science {\bf 303}, 1490 (2004).
\bibitem{senthil2} T.~Senthil, L.~Balents, S.~Sachdev, A.~Vishwanath,
and M.~P.~A.~Fisher, Phys. Rev. B  {\bf 70}, 144407 (2004).
\bibitem{RKMMSS08} R.~K.~Kaul, M.~A.~Metlitski, S.~Sachdev and C.~Xu, Phys. Rev. B {\bf 78}, 045110 (2008).
\bibitem{TS08} T. Senthil, Phys. Rev. B {\bf 78}, 045109 (2008).
\bibitem{SS88} B.I. Shraiman and E.D. Siggia, Phys. Rev. Lett. {\bf 61}, 467 (1988).
\bibitem{blackburn13} E. Blackburn {\it et al.\/}, Phys. Rev. Lett. {\bf 110}, 137004 (2013).
\bibitem{Yang11} H.-B. Yang, J. D. Rameau, Z.-H. Pan, G. D. Gu, P. D. Johnson, H. Claus, D. G. Hinks, and T. E. Kidd, Phys. Rev. Lett. {\bf 107}, 047003 (2011).
\bibitem{LT07a} N. Doiron-Leyraud, C. Proust, D. LeBoeuf, J. Levallois, J-B. Bonnemaison, R. Liang, D. A. Bonn, W. N. Hardy and L. Taillefer Nature {\bf 447}, 565-568 (2007).
\bibitem{LT07b} D. LeBoeuf {\it et al.\/}, Nature {\bf 450}, 533-536 (2007).
\bibitem{LT09} L. Taillefer, J. Phys.: Condens. Matter {\bf 21}, 164212 (2009).
\bibitem{NHSS11} N. Harrison and S. E. Sebastian, Phys. Rev. Lett. {\bf 106}, 226402 (2011).
\bibitem{NHSS12} S. E. Sebastian, N. Harrison and G. G. Lonzarich, Rep. Prog. Phys. {\bf 75}, 102501 (2012).
\bibitem{MG13} N. Barisic {\it et al.\/}, Nature Physics {\bf 9}, 761-764 (2013).
\bibitem{SS14} S. E. Sebastian, 	N. Harrison, F. F. Balakirev,	M. M. Altarawneh, P. A. Goddard, Ruixing Liang, D. A. Bonn, W. N. Hardy and G. G. Lonzarich, Nature {\bf 511}, 61-64 (2014).
\bibitem{LT14} N. Doiron-Leyraud {\it et al.}, arXiv:1409.2788.
\bibitem{GB} S.C. Riggs, O. Vafek, J. B. Kemper, J.B. Betts, A. Migliori, W. N. Hardy, Ruixing Liang, D. A. Bonn and G.S. Boebinger, Nature Physics {\bf 7}, 332-335 (2011). 
\bibitem{KS} S. Kawasaki, C. Lin, P. L. Kuhns, A. P. Reyes, and G-Q Zheng, Phys. Rev. Lett.{\bf 105}, 137002 (2010).
\bibitem{AADCSS} A. Allais, D. Chowdhury and S. Sachdev, Nature Communications {\bf 5}, 5771 (2014).
\end{thebibliography}
\end{document}